# Probing Intrinsic Material Conductivity in Two-Terminal Devices: A Resistance-Difference Method

Yang Lu, and I-Wei Chen [a]

Department of Materials Science and Engineering, University of Pennsylvania, Philadelphia, PA 19104-6272, USA

**Abstract**

It is generally impossible to separately measure the resistance of the functional component (i.e., the intrinsic device materials) and the parasitic component (i.e., terminals, interfaces and serial loads) in a two-terminal device. Yet such knowledge is important for understanding device physics and designing device systems. Here, we consider a case where an electric current, temperature, or magnetic field causes a small but identical relative conductivity change $\Delta\sigma/\sigma$ of the device materials. We find an exact solution to this relative change by a simple resistance-data analysis of similarly configured two-terminal devices. The solution is obtainable even if the change is quite small, say, less than 0.1%. In special cases of small relative changes in parasitic resistance, the absolute parasitic resistance is also obtainable. Our method is especially useful for studying the switching and transport characteristics of the emergent non-volatile resistance memory.

---

[a] Electronic mail: iweichen@seas.upenn.edu

The functional device component in an electronic device always coexists with a parasitic component,[1-5] which includes the device terminals, interfaces between the functional component and the terminals, leakage pathways and serial loads. So the measured device response (e.g., voltage) inevitably comprises the response of the parasitic component, which masks the true response of the device material.[6-8] The four-point/terminal method can isolate the signal of the device material and is the method of choice, for example, for measuring the sheet resistivity of a semiconductor thin film.[9] But the electrode distance is often too short to implement the four-point method in two-terminal devices, e.g., diodes—rectifier diodes and light-emitting diodes,[10] solar cells,[6] and non-volatile memories—phase change memory (PCRAM)[11] or resistance memory (RRAM)[3-4,8]. While in some configurations, a local probe such as scanning voltage probe may help measure the contact resistance directly,[12] in most cases it is not applicable since two-terminal devices usually have more than one buried interface. In this respect, one of the most measurement-challenging configurations is a three-film stack, with a device material film sandwiched between a top electrode film and a bottom electrode film, all three films having a large aspect ratio—the ratio of the lateral dimension to the thickness—that often exceeds $10^2$ or even $10^4$. Nevertheless, it is important to interrogate the device-material, in particular its resistance, which may dictate the device performance when a current flows from one electrode to the other.

In this paper, we present a method to separate the resistance of the functional component from that of the parasitic component for a set of devices that comprises *self-similar* device materials and employs identical terminals, device structures and serial loads. Here, self-similarity refers to a similar voltage response—to the stimuli of current ($I$), temperature ($T$), magnetic field ($H$), etc.—that differs by at most a multiplication constant. Equivalently, self-similarity is manifest if all the device materials in different devices experience the same fractional change of resistance or conductivity ($\Delta\sigma/\sigma = -\Delta R/R$) in response to a stimulus. For device materials, such self-similarity is quite common as it is often rooted in material physics. For example, the material conductivity/resistivity may follow a power law, $T^p$, where $p$ is an exponent of the order of unity,[13] or its magentoresistance may obey a power law such as $H^2$ or $H^{1/2}$.[14] For devices that contain such materials, our method can be applied to analyze their behavior. Another application is when such self-similarity or power law is in doubt; our method can provide a robust test to see whether self-similarity or power law actually holds or not. The method is especially valuable and timely for non-volatile resistance memory (also known as RRAM), which is an emergent two-terminal device.[15] Serially connected to the load resistance of electrodes, interface and external lines,[16-17] the resistance of a memristive material can switch, under a voltage, between a high resistance state (HRS) and a low resistance state (LRS), often with additional intermediate states between the two.[17-19] Applying our method, one can (a) determine whether different resistance states (which may contain, for example, multiple conducting filaments[20-21]) are self-similar or not, and if they are, (b) interrogate the switching and transport characteristics of the memristive material.

To extract the self-similar response, e.g., in the form of $\Delta\sigma/\sigma = -\Delta R/R$, we used a method of resistance difference. Specifically, we analyzed the difference in the device resistances of a pair of self-similar devices sharing a common parasitic component. The method takes advantage of the fact that the difference resistance has nothing to do with the parasitic resistance, which cancels out; as such, it is just the resistance difference of the two device materials. Therefore, if self-similarity holds for the device materials, it must also manifest in the difference resistances of their device pairs. The converse is true too: Self-similarity of the difference resistances of device pairs implies self-similarity of their device materials.

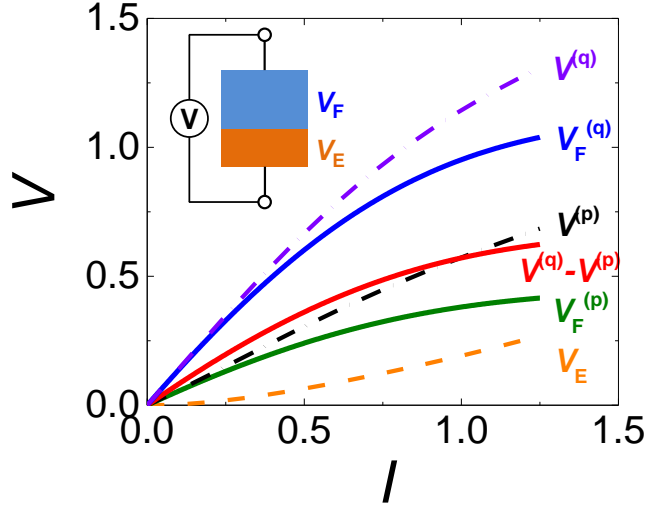

FIG 1. Simulated *I-V* curves of two two-terminal devices: *p* and *q*, with total voltage *V* being the sum of voltage of self-similar device material ($V_F^{(q)}=2.5V_F^{(p)}$) and voltage of identical terminal resistance ($V_E$). While the shapes of $V^{(q)}$ and $V^{(p)}$ are different, $V^{(q)}-V^{(p)}$ shows the same *I-V* trend as $V_F^{(p)}$, only differing by a multiplication factor=1.5; i.e., $V^{(q)}-V^{(p)}$, $V_F^{(p)}$ and $V_F^{(q)}$ are all self-similar. Inset: Schematic of a two-terminal device with two components.

This idea is illustrated in **Fig. 1**, in which a device is shown in the inset and is conceptually separated into two serial parts, (a) the device material carrying a voltage $V_F$, so named because it quite often involves a film, and (b) the load carrying a voltage $V_E$, so named because it always involves electrodes in addition to lead wires, spreading resistance and contact resistance—between the electrodes and the device material. The voltage response to a current is drawn for two self-similar devices, $V^{(p)}$ and $V^{(q)}$. Here, $V^{(p/q)} = V_F^{(p/q)} + V_E$, where $V_F^{(p/q)}$ is the *I-V* curve of the film material in the (*p/q*)-device, which is self-similar, and $V_E$ is the *I-V* curve of the parasitic part, which is the same for (*p*) and (*q*). Note that $V^{(p)}$ may look quite different from $V^{(q)}$—in **Fig. 1**, $V^{(p)}$ appears as a straight line but $V^{(q)}$ is concave downward, even though $V_F^{(p)}$ and $V_F^{(q)}$ only differ by a multiplication constant. Mathematically, instead of finding $V_F^{(p)}$ and $V_F^{(q)}$, we will find $V^{(q)} - V^{(p)}$, which differs from $V_F^{(p)}$ and $V_F^{(q)}$ by a multiplication constant, i.e., they are all self-similar. For most purposes, this knowledge of $V_F$ is enough to inform the device physics. Importantly, it can be obtained without knowing $V_E$.

In the following, the devices that have the same load in contact with the self-similar device (film) materials are referred to as self-similar devices of the same configuration. These devices form a device set to be analyzed below. Here, the self-similar device (film) materials may have the same composition and microstructure but of different fabricated thickness, or they may be the same film but are set in different resistance states. In an RRAM that have multiple HRS resistance states or multiple LRS states,[17-19] a device set may be constructed by including several HRS states or several LRS states. (But we do not advice mixing HRS states and LRS states because they are usually distinct and unlikely to behave in a self-similar way.[22-23]) To describe self-similarity, we use a multiplication factor $\phi$, which we let, without loss of generality, $V_F(I, \phi) = (1/\phi) V_F^*(I)$, where $V_F^*$ is the (device-material) film voltage of a reference device. This reference device need not actually exist in the above device set; even if it does, its choice need not be unique.

We now write the device response as

$$V(I,\phi) = V_F(I,\phi) + V_E(I) = \frac{1}{\phi} V_F^*(I) + V_E(I) \qquad (1)$$

Next, we compare any two devices, $p$ and $q$, in the set to obtain the following "difference voltage" that is proportional to $V_F^*$

$$V^{(q)}(I,\phi^{(q)}) - V^{(p)}(I,\phi^{(p)}) = (\frac{1}{\phi^{(q)}} - \frac{1}{\phi^{(p)}})V_F^*(I) \tag{2}$$

The above is true in both current directions even though only the positive direction is shown in **Fig. 1**. To include the effect of other perturbations, we consider the voltage response to any "field" $J$, such as temperature and magnetic field. Including both $I$ and $J$ while still assuming self-similarity, we generalize Eq. (1-2) to obtain

$$V(I,J,\phi) = V_F(I,J,\phi) + V_E(I,J) = \frac{1}{\phi}V_F^*(I,J) + V_E(I,J) \tag{3}$$

$$V^{(q)}(I,J,\phi^{(q)}) - V^{(p)}(I,J,\phi^{(p)}) = (\frac{1}{\phi^{(q)}} - \frac{1}{\phi^{(p)}})V_F^*(I,J) \tag{4}$$

For applications such as resistance memory, it is also convenient to recast the above results into resistance, defined as $R=V/I$. With the additional definitions $R_E(I,J)=V_E(I,J)/I$, $R_F(I,J,\phi)=V_F(I,J,\phi)/I$, and $R_F^*(I,J,\phi)=V_F^*(I,J,\phi)/I$, we rewrite Eq. (1) as

$$R(I,J,\phi) = \frac{1}{\phi}R_F^*(I,J) + R_E(I,J) \tag{5}$$

We now contemplate the effect of two perturbations. (a) The resistance change due to temperature is measured to determine the temperature coefficient of resistance. (b) The resistance change due to a magnetic field is measured to determine magnetoresistance. In both cases, one deals with the resistance variation $\Delta R$ that arises when the current and field change from $(I_0, J_0)$ to $(I, J)$. Introducing the following definitions

$$\Delta R(I,J,\phi) = R(I,J,\phi) - R(I_0, J_0, \phi) \tag{6}$$

$$\Delta R_E(I,J) = R_E(I,J) - R_E(I_0,J_0) \tag{7}$$

$$\Delta R_F^*(I,J) = R_F^*(I,J) - R_F^*(I_0,J_0) \tag{8}$$

$$\Delta R_F(I,J,\phi) = R_F(I,J,\phi) - R_F(I_0,J_0,\phi) = \frac{1}{\phi}\Delta R_F^*(I,J) \tag{9}$$

we get, from Eq. (5),

$$\Delta R(I,J,\phi) = \frac{1}{\phi}\Delta R_F^*(I,J) + \Delta R_E(I,J) \tag{10}$$

For any device/state pair (*p,q*), we find the following equation to hold

$$\frac{\Delta R^{(q)}(I,J,\phi^{(q)}) - \Delta R^{(p)}(I,J,\phi^{(p)})}{R^{(q)}(I,J,\phi^{(q)}) - R^{(p)}(I,J,\phi^{(p)})} = \frac{\Delta R_F^*(I,J)}{R_F^*(I,J)} \tag{11}$$

This is a remarkable result in that the right-hand side, which contains neither $R_E$ nor $\phi$, and is thus an intrinsic property of the device material, can be obtained from the left-hand side that contains only experimentally measurable quantities. The above intrinsic property can be alternatively represented in terms of conductivity σ. To do this, we write $R_F^*$ as $G/\sigma_F^*$, where $G$ is a geometric constant (e.g., the ratio of thickness to area in the case of a film), and $\sigma_F^*$ is the conductivity of the reference (device-material) film. With this definition, Eq. (11) reduces to

$$\frac{\Delta R^{(q)}(I,J,\phi^{(q)}) - \Delta R^{(p)}(I,J,\phi^{(p)})}{R^{(q)}(I,J,\phi^{(q)}) - R^{(p)}(I,J,\phi^{(p)})} = -\frac{\sigma_F^*(I,J) - \sigma_F^*(I_0,J_0)}{\sigma_F^*(I_0,J_0)} \equiv -\frac{\Delta \sigma_F^*}{\sigma_F^*(I_0,J_0)} \equiv -x(I,J) \tag{12}$$

with $\sigma_F^*(I,J) \equiv [1+x(I,J)]\sigma_F^*(I_0,J_0)$ and $x(I_0,J_0)=0$.

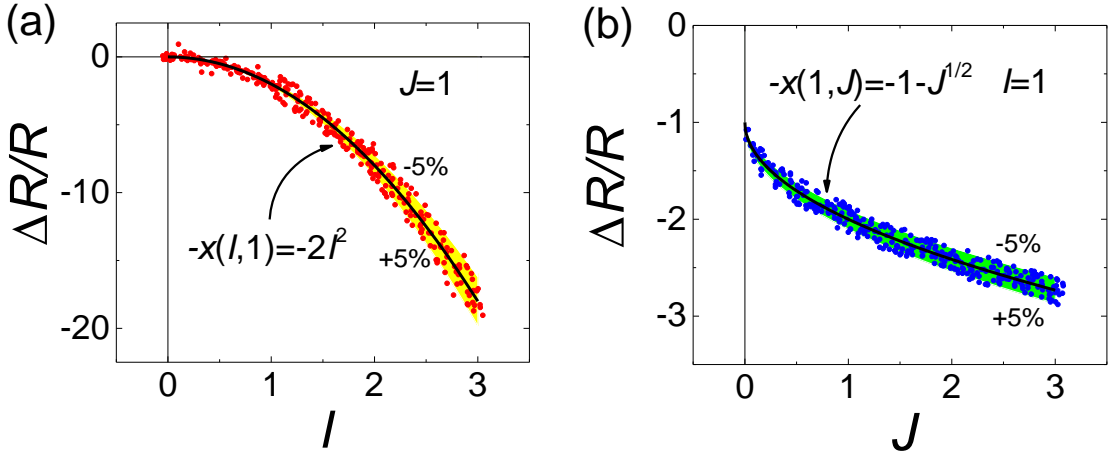

FIG.2. Simulated $\Delta R/R$ as a function of: (a) $I$, at fixed $J=1$ and (b) $J$, at fixed $I=1$, using $(I_0, J_0)=(0,0)$, $R_F^*(0,0)=10$, $R_E(0,0)=10$, $x(I,J)=I^2(1+J^{1/2})$, $\Delta R_E(I,J)=I^{1/2}(1+J^2)$+random noise. As predicted by Eq. (12), most $\Delta R/R$ data cluster around $-x(I,J)$, within 5% (shown as colored bands), and the statistical variations are relatively similar in both small and large $I$ (or $J$) regions.

The above set of results provides both the solution to the problem and the self-consistency check of our assumption. (i) If self-similarity holds, the intrinsic property being a universal function of $(I,J)$ can be obtained from any device/state pair $(p,q)$, and the same result should always be obtained. (ii) Conversely, if the right-hand side obtained from different device/state pairs does differ, then the self-similarity assumption must be invalid. This is demonstrated in **Fig. 2** which simulates self-similar devices/states with multiple $\phi$ values but of identical configuration: In **Fig. 2a**, we set $J=1$, and in Fig. 2b, we set $I=1$. In both simulations, $x(I,J)$ is chosen to cover a range of $\Delta R/R$ (which depends on the magnitude of $I$ and $J$), and realistic random noise is introduced to the $\Delta R_E(I,J)$ "data" throughout the range of $I$ covered in **Fig. 2a**, and $J$ covered in **Fig. 2b**. They illustrate that the right-hand side of Eq. (11) statistically agrees with the predicted $-x(I,J)$ of Eq. (12) within 5% (shown by the colored bands). In this way, they (i) validate the self-similarity assumption, a posteriori, and (ii) determine the intrinsic property $x(I,J)$ of the device materials. Note that the statistics of the simulated results depends only on the ratio of the noise to $\Delta R_E(I,J)$ and is independent of the magnitude of $\Delta R(I,J)$ and $x(I,J)$. Therefore, even a very small $\Delta R_F/R_F$ (or $x(I,J)$) can be determined using Eq. (11-12), which is exact. (We have analyzed data of an LRS RRAM and obtained $\Delta R_F/R_F$ ranging from 0.1% to 10%, with a $(T,B)$ dependence in excellent agreement with the theoretical prediction of mesoscopic physics.[13]) This makes our solution especially valuable for studying various LRS states in an RRAM because their resistance values may be of the same order of magnitude as, or even smaller than, $R_E$.[17,23-24]

While Eq. (11-12) were derived for a device/state pair $(p,q)$, we could also determine the intrinsic property $\Delta\sigma_F^*/\sigma_F^*$ from a device/state quartet $(r,s,t,v)$. Conceptually, this involves two levels of difference-resistance computation, in the following sequence: First combine the $(r,s)$ pair to make another "$p$", and likewise combine the $(t, v)$ pair to make another "$q$"; next use the new pair $(p,q)$ and Eq. (11-12) to obtain $\Delta R_F^*/R_F^*$ or $\Delta\sigma_F^*/\sigma_F^*$. The procedure works because it serves to cancel $R_E$ and to eliminate $\phi$ in the same way as before. Since higher order multiplets such as octates, etc., can also be similarly used, we can generalize the above procedure for $N$ self-similar devices/states of identical configurations. There are altogether $C(N,2)+C(N,4)+\ldots\ldots+C(N,N-2n)$ sets of resistance (differences) to consider. Here, $C$ is the binomial coefficient and the maximum $2n$ is either $N-1$ or $N$.

The above solution for $\Delta R_F^*/R_F^*$ and $\Delta\sigma_F^*/\sigma_F^*$ is exact. Below we will provide an approximate solution for $R_E$, which is possible if $\Delta R_E$ is small. From Eq. (5) and Eq. (13), we have

$$R(I,J,\phi) = \frac{1}{\phi}R_F^*(I,J) + R_E(I,J) = \frac{1}{\phi}\frac{G}{[1+x(I,J)]\sigma_F^*(I_0,J_0)} + R_E(I,J) \quad (14)$$

From this, we obtain

$$\Delta R(I,J,\phi) = -\frac{1}{\phi}\frac{x(I,J)}{1+x(I,J)}\frac{G}{\sigma_F^*(I_0,J_0)} + \Delta R_E(I,J) \qquad (15)$$

Combining Eq. (14-15) to eliminate $(1/\phi)G$, we obtain

$$R(I,J,\phi) = -\frac{\Delta R(I,J,\phi) - \Delta R_E(I,J)}{x(I,J)} + R_E(I,J) \qquad (16)$$

Since $x(I,J)$ is known from Eq. (13), we can plot $R$ against $\Delta R/x$ from devices/states of different $\phi$ but at the same $(I,J)$ to determine $R_E(I,J)$ from the intercept. This is illustrated in **Fig. 3** using simulated data for an arbitrarily chosen $x(I,J)$ and $\Delta R_E(1,1)=0.01\Delta R_F^*(1,1)+$random noise. As expected from Eq. (16), all the data points statistically cluster around a straight line with a slope of negative unity, within 5% (shown as the yellow band), and its intercept $R_E = 9.9$ is very close to the prescribed $R_E(I,J) = 9.93$, because $\Delta R_E(I,J)/x$ in Eq. (16) is very small.

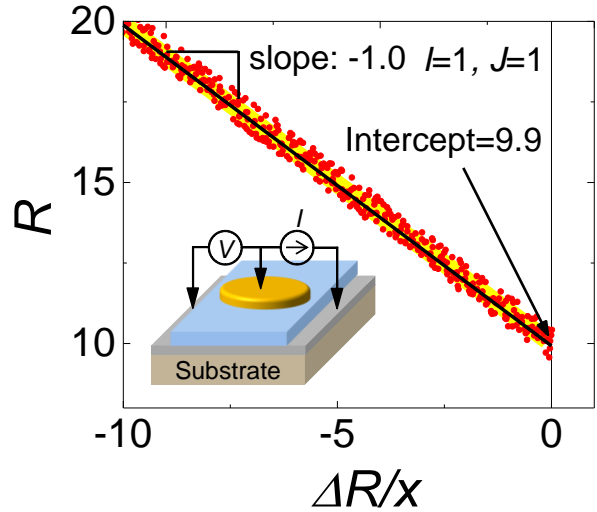

FIG.3. $R(I, J, \phi)$ against $\Delta R/x$, simulated from devices/states of different $\phi$ but at same $(I,J)=(1,1)$. $(I_0, J_0)=(0,0)$, $R_F^*(0,0)=R_E(0,0)=10$, $x(I,J)=I^2(1+J^{1/2})=3$, and $\Delta R_E(1,1)=0.01\Delta R_F^*(1,1)+$random noise. As predicted by Eq. (16), most data cluster around a straight line of a slope of negative unity, within 5% (shown as yellow band), and the statistical variations are essentially the same everywhere. The intercept resistance is very close to the prescribed $R_E(I,J)=9.93$, since $\Delta R_E(I,J)/x(I,J)/=0.02$ is very small, at $(I,J)=(1,1)$. Inset: Schematic of three-point measurement configuration for a two-terminal device, with patterned top electrode (yellow) and continuous bottom electrode (gray).

Obviously, the above procedure works best with a larger $x$, which may be realized under a strong field, $J \gg 1$, e.g., a strong magnetic field. In such case, the following procedure can actually determine $R_E(I,J)$ for all $J$. First, we measure $R(I,J,\phi)$ for $J \gg 1$ to take advantage of the large $x$, which allows us to use Eq. (16) to find $R_E(I,J)$. After subtracting $R_E(I,J)$ from $R(I,J,\phi)$, we obtain $R_F(I,J,\phi)$ for all the devices in this favorable case. Next, we recall that $R_F(I,J,\phi)$ is proportional to $\phi^{-1}$. Therefore, referring to any device in the set as the reference state with $\phi^*$, we can express the resistance of other devices/states as $R_F(I,J,\phi)=R_F(I,J,\phi^*)(\phi^*/\phi)$. From this, we obtain $\phi^*/\phi$ for all the devices, still under the same $(I,J)$. Third, for any other $J$ that may not satisfy $J \gg 1$, thus may not have a large $x$, we still have the following relation because we have not changed $\phi$

$$R(I,J,\phi) = R_F^*(\phi^*/\phi) + R_E(I,J) \tag{17}$$

Here $R_F^* = R_F(I,J,\phi^*)$ denotes the film resistance of the reference device/state. Fourth, plotting the measured device resistance $R(I,J,\phi)$ against $\phi^*/\phi$, which we already obtained for all the devices/states from the $J \gg 1$ data, we should obtain a straight line from which the intercept gives $R_E(I,J)$. Fifth, having determined $R_E(I,J)$ for all $J$, we can subtract it from the device resistance $R(I,J,\phi)$ to finally have $R_F(I,J,\phi)$ for all $J$, for all the devices/states. The problem of film resistance of the two-terminal devices is now completely solved.

Obviously, an improvement in measurement configuration will help, as illustrated by the following example. In some devices, the device material occupies a small footprint but part of the electrode extends to a large area. Such electrode often has a large spreading resistance, which may provide a large $\Delta R_E(I,J)$ thus rendering Eq. (16) useless. A remedy is to employ the three-point method illustrated in the inset of **Fig. 3**, which provides results that are free of the spreading resistance. The remaining $R_E$ contains neither the spreading resistance nor the bottom-electrode resistance, but it does include the top-electrode resistance; yet overall it is more likely to give a smaller $\Delta R_E$. The method is especially advantageous in magnetic measurement because the spreading resistance, associated with a large area, could experience more magnetoresistance than the device-material film does.

In conclusion, when an electric current, temperature, or magnetic field causes a small but identical relative conductivity change $\Delta\sigma/\sigma$ of the device material in a set of similarly configured two-terminal devices, an exact solution to this change can be obtained by a simple analysis of the two-terminal resistance data. This solution is useful for interrogating the self-similarity of device materials, such as the resistances of filaments in different resistance states in RRAM, and their power-law dependence on $T$ and $H$ expected from mesoscopic physics. The solution can therefore help acquire intrinsic device data and elucidate device physics for all types of two-terminal electronic devices.

This research was supported by the US National Science Foundation Grant No. DMR-1409114. The use of facilities at Penn's LRSM supported by DMR-1120901 is gratefully acknowledged.